\documentclass[10pt,conference,letterpaper]{IMSTemplate}

\usepackage{times,amsmath,epsfig}

\usepackage{todonotes} % todo remover daqui
\usepackage{algorithm,algpseudocode}
\usepackage{url}
\usepackage{graphicx}

\title{Genetic Soundtracks: Creative Matching of Audio to Video}

\author{
Jorge Gomes{\small $~^{\#}$}, Fernando Silva{\small $~^{\#}$} and Teresa Chambel{\small $~^{*}$}
\vspace{12pt}\\
$~^{\#}$LabMAg, Faculty of Sciences, University of Lisbon, 1749-016 Lisbon, Portugal\\
$~^{*}$LaSIGE, Faculty of Sciences, University of Lisbon, 1749-016 Lisbon, Portugal
}

\begin{document}
\maketitle

%****************************************************************************
%
\begin{abstract} 
The matching of the soundtrack in a movie or a video can have an enormous influence in the message being conveyed and its impact, in the sense of involvement and engagement, and ultimately in their aesthetic and entertainment qualities. Art is often associated with creativity, implying the presence of inspiration, originality and appropriateness. Evolutionary systems provides us with the novelty, showing us new and subtly different solutions in every generation, possibly stimulating the creativity of the human using the system.
In this paper, we present Genetic Soundtracks, an evolutionary approach to the creative matching of audio to a video. It analyzes both media to extract features based on their content, and adopts genetic algorithms, with the purpose of truncating, combining and adjusting audio clips, to align and match them with the video scenes.
\end{abstract}

%****************************************************************************
% NOTE keywords are not used for conference papers so do not populate
% them

%****************************************************************************
\section{Introduction}

The amount of home-recorded videos is dramatically increasing due to the extensive use of personal video cameras and recording tools. Without post-production, these videos typically appear very amateur and raw. One of the most common processes consists on creating a soundtrack for a given video. This task, however, is usually very time consuming, since it involves manually choosing and editing the audio snippets that the user considers most appropriate for each video segment. With the increasing amount of accessible video clips and movies over the internet, the possibility of exploring alternate and creative editions in this context is an interesting subject. After all, the soundtrack can considerably affect the impact, appeal and engagement of the videos.

A system that uses evolutionary algorithms can help in the creative process of choosing the adequate soundtrack for a video. An evolutionary algorithm is a non-deterministic optimization method that in each run can generate different solutions, even if the initial parameters are the same. Although the solution generation process is intended to be automatic, there are some characteristics of the soundtrack that are very hard to analyse, such as the emotions and lyrics of the music. Evolutionary algorithms can provide an answer to this problem, as the user can choose among a wide range of (apparently) viable solutions provided by the algorithm.

In this paper, we present an automated method for the matching of audio (mainly music) to videos, given an arbitrary set of audio files, and a source video selected by the user. The presented method starts by automatically extracting characteristics from the video and audio files. In a second phase, the system uses a genetic algorithm to match the audio snippets to video, using the media characteristics to obtain the notion of fitness. We developed a tool using Java and Processing\footnote{\url{http://processing.org}} to implement and test the proposed method.

Automatic video and audio content analysis is performed to extract characteristics from both media: measures through time and instants of significant changes, providing the basis for the matching process. The matching process is based on genetic algorithms (GAs), where each chromosome represents a solution -- a sequence of audio snippets (or silences) with the same duration of the video. To fit the audio tracks to the video, the genetic algorithm can select, truncate, combine and discard audio clips, therefore creating a soundtrack that, at each moment, is adequate to the current video segment. This produces a pleasurable result with very little effort by the user.

The notion of soundtrack-video adequacy tries to mimic the subjective notion of fitness that users consider in manual matching. The adequacy is defined as a correspondence function between the extracted characteristics from audio and video segments, and can be customised by the users to produce a result that meets their requirements.

\section{Related Work}

An adequate correspondence between audio and video channels is widely recognised as an important element in the movie visualisation experience. For example, Grimes~\cite{grimes1990} conducted an empirical study where it was demonstrated that the audio-video correspondence plays an important role in attention and memory. Studies have also shown that poor sound quality degrades the perceived video image quality~\cite{neuman}, strengthening the notion that audio and video have a strong connection in the visualisation experience.

In Synesthetic Video~\cite{chambel2010seeingcolors}, the authors explored the relation of visual and auditory properties to experience video in cross-sensorial modes, resulting in ways to hear its colours (synthesised, not matching of existing audio) and to influence its visual properties with sound and music, through user interaction or ambient influence.
The motivations behind this work were accessibility, enriching users' experiences, and stimulating and supporting creativity. In~\cite{foote02creatingmusic} is performed automatic and semi-automatic selection and alignment of video segments to music. The objective of the proposed method is to create a suitable video track for the given soundtrack, which is the opposite of our work. The process is based on the detection of audio and video changes, plus camera motion and exposure, to help determine suitability between the video and audio tracks. Deterministic methods are proposed for the alignment of audio and video, such as best-first search.

Evolutionary computation has been widely used in art domains, such as music generation~\cite{romero} and video generation~\cite{videoArtEA}. In~\cite{hua04automaticgeneration}, music videos are automatically generated from personal home videos, based on the extraction and matching of temporal structures of video and music, using genetic algorithms to find global optimal solutions. These solutions may involve repetitive patterns in video based on those found in the music. In MovieGene~\cite{moviegene} the authors used genetic algorithms to explore creative editing and production of videos, with the main focus on visual and semantic properties, by defining criteria or interactively performing selections in the evolving population of video clips, which could be explored and discovered through emergent narratives and aesthetics.

\section{Media Features Extraction}

Our first step in matching video and audio is the automatic extraction of features from the media. The features are extracted only once for each medium, when they are first added to the system. Many features can be extracted from both audio and video, in order to possibly exploit synesthetic relations~\cite{chambel2010seeingcolors}, such as audio frequency, levels or rhythm, and video colour, lightness or movement. The objective is to extract from both audio and video characteristics that are perceptually relevant to the user.
To demonstrate the concept, we chose to base the matching process in these two easily extractable features: video movement and audio levels. These two features are intuitively related. In cinema, for example, fast moving and action-packed scenes are typically associated with a loud and vivid soundtrack, while slow moving scenes are usually accompanied by a soft soundtrack.

\subsection{Video Analysis}

To extract the video movement we use frame differencing between every adjacent frame in the video~\cite{Zhang1993}. The frame differencing values are processed to obtain two metrics: 1) the instants in time where scene cuts occur; and 2) the evolution of the average video movement along time. To obtain the scene cuts, each frame differencing value is compared with the average frame difference of the last $N$ frames (correspondent to 1,5\,s). If the value is greater than the average of the last by a threshold $T$, then it is considered a scene cut. To retrieve the average video movement through time, video frames are partitioned into blocks of 500\,ms each, and the average movement inside each block of frames is calculated. An example of a video segment analysis is depicted in Figure~\ref{fig:result} (\emph{Video}).

\subsection{Audio Analysis}

As in the video analysis, the objective is to collect information about the average audio levels over time and to identify instants where significant changes in the music occur. The audio samples are obtained at a rate of 10\,ms and are analysed with a succession of Fast Fourier Transforms (FFT) over time. The audio level is extracted and the frequency components of each sample are aggregated and averaged in 10 bands, one corresponding to each octave, ranging from about 21 Hz to 22 KHz. The extracted characteristics of the samples are then averaged in blocks of 500\,ms, in order to obtain more meaningful data and easier to process.

The frequency bands are used to identify significant changes in the music. These changes are captured by two methods: i) identification of significant changes in the bands values, when compared to the average values of the current music segment; ii) identification of significant changes in the bands values variation (measured by the standard deviation of the bands values). The threshold that defines what is a significant change is key in the process, and is calculated dynamically according to the following factors:

\begin{itemize}
\item The audio level of the sample being analysed. Higher level values are perceptually less distinct and so the threshold increases linearly with the level.
\item The time passed since the last music segment cut. It is not desirable that a music segment is too short or too long, so the threshold is higher when the sample being analysed is too close to the last segment cut.
\item Pre-defined threshold scaling factor that determines whether the algorithm should be more or less sensitive to changes in music.
\end{itemize}

This process allow us to obtain music segments that are perceptually identified by the listener. An example of two musics segmented with this method is depicted in Figure~\ref{fig:result} (\emph{Music 1 - Bittersweet} and \emph{Music 2 - Fuel}).

\section{Matching Audio to Video}

\subsection{Genetic Representation and Initialisation}
The matching process between audio clips and video scenes is modelled as an optimisation task performed by Genetic Algorithms (GAs)~\cite{mitchellGA}.
The information regarding each 500 milliseconds audio sample is genetically encoded as a \emph{gene}.
Genes maintain extracted information such as the sample level values and sample position in the corresponding audio clip.
Each string of genes is denoted as a \emph{chromosome}, an individual of the population, which constitute a possible solution to the matching process.
Chromosomes have a pre-determined size corresponding to the number of genes necessary to keep up with the entire video.
Adjacent samples belonging to the same audio snippet are placed sequentially in the chromosomes therefore maintaining its original sequence.

The initial population of chromosomes is generated through a semi-random process. Each chromosome is generated by putting together randomly chosen segements of audio clips and silences.
In Algorithm~\ref{alg:initialisationAlgorithm}, we summarise the method used for the creation of each chromosome in the initial population. 

\begin{algorithm}
\caption{Algorithm for generating each chromosome in the initial population.}
\label{alg:initialisationAlgorithm}
\begin{algorithmic}
\State $remaining \gets chromosomeSize$ 
\While{$remaining > 0$}
	\State $option \gets$ choose $audioclip$ or $silence$
	\If{$option = audioclip$}
        \State $clip \gets$ select one of the available musics
        \State $startPoint \gets$ select one of the music's cuts
        \State $endPoint \gets$ random($startPoint,musicLength$)
        \State $endPoint \gets$ min$(endPoint,chromosomeSize)$
        \State $snippet \gets$ truncate($clip$, $startPoint$, $endPoint$)
    \Else
    	\State $length <- $ random(1,$remaining$)
    	\State $snippet <-$ silence($length$) 
    \EndIf
    \State add $snippet$ to chromosome
    \State $remaining \gets remaining +$ length($snippet$)
\EndWhile
\end{algorithmic}
\end{algorithm}

\subsection{Genetic Operators and Fitness Function}
As in the canonical GA, at each generation a subset of the population is selected to create a new population.
Chromosomes are evaluated, selected and submitted to distinct genetic operators.

The \emph{fitness function} evaluates the adequacy between the video and soundtracks based on the weighted composition of several factors.
Weights are customisable and therefore the user may regulate the importance of the criteria involved.
Defined criteria include:
\begin{itemize}
\item Correlation between audio levels at a certain time and the video movement of the contemplated scene, which should be maximised.
\item Average length and number of audio snippets, which favours snippets of larger duration in order to avoid excessive changes between audio clips.
\item Quantity of silence in the audio snippets, which should not be too much.
\item Temporal matching between the audio sequence and the video, which tries to avoid, for instance, that an audio clip starts in the middle of a scene.
\end{itemize}
After the evaluation process, a percentage of the chromosomes are selected through \emph{tournament selection}, to originate a new population.

\begin{figure}[tb]
\centering
\includegraphics[scale=1]{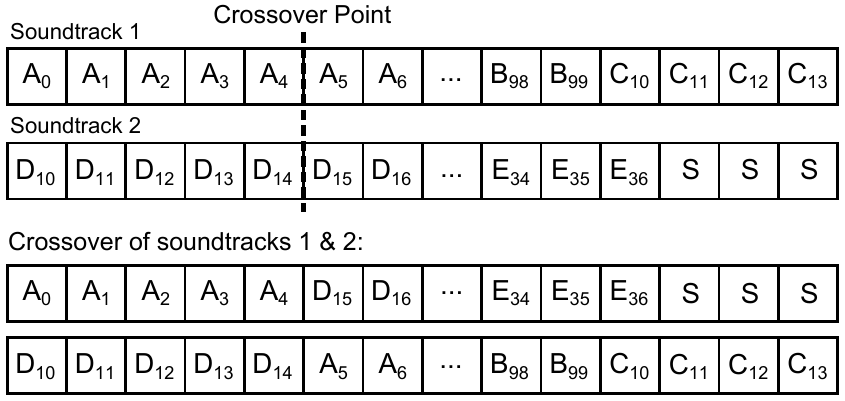}
\caption{Example of the crossover operator. A crossover point is chosen afterwhich the two sequences are recombined, therefore producing two new ones. $X_{i}$ means audio frame $i$ of the music $X$.}.
\label{fig:crossover}
\end{figure}

\emph{Reproduction} is the process by which a new set of solutions is generated, where pairs of chromosomes are combined to originate two new ones.
New chromosomes are created by randomly choosing a crossover point, an instant in time that splits each of the parent audio sequences.
The first part of each sequence is attached to the last part of the other sequence thus generating two new sequences.
In Fig.~\ref{fig:crossover}, we illustrate the crossover process.

\emph{Mutations} introduce changes in the chromosomes of the population.
Each audio snippet in the chromosome is mutated with a given probability.
If a snippet is to be mutated, mutation functions may: (i)~alter the position of an audio snippet by moving it in the sequence, (ii)~replace a snippet by a different one, which may or not exist be contemplated in the chromosome, (iii)~stretching or shrinking a certain snippet, or (iv)~remove an audio snippet thus creating silence.

\emph{Decimation} operator removes from the population chromosomes that violate the system's restrictions.
In order to avoid radical audio changes, each audio snippet composing in a chromosome must have a minimum duration.
This way, we avoid a large fragmentation of the candidate solutions and accelerate convergence to a good solution by imposing \emph{minimal criteria}.
The second restriction is related to the starting point of each audio segment.
The audio segments in a chromosome should start in a transition point of the corresponding music, in order to avoid very rough transitions in the soundtrack.

\begin{figure}[tb]
\centering
\includegraphics[scale=1.1]{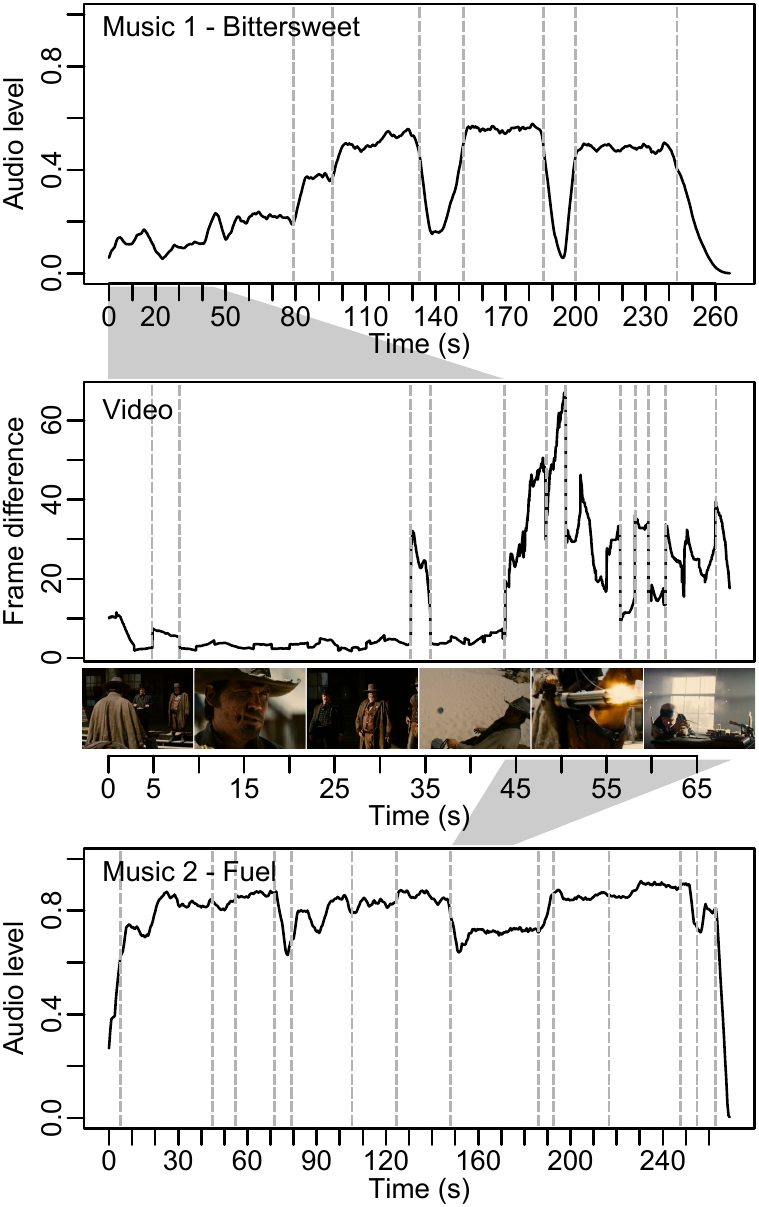}
\caption{The matching of two songs (Bittersweet and Fuel) and a video.
The first and third graphics show the average audio levels through time, and the significant changes detected.
The second graphic shows the movement of the video and scene changes detected.
Gray areas represent the segment of audio selected to keep up with a given set of video scenes.}
\label{fig:result}
\end{figure}

\section{Results}

A new technology is largely justified by useful applications and valid results.
In order to validate Genetic Soundtracks, we conducted tests with a large and varied set of complete songs, and excerpts of video with distinct lengths.
The measures extracted, audio level and frame differencing, are not always able to capture the intrinsic features of video scenes and audio clips.
Nonetheless, the majority of the results were promising, being both compelling and pleasurable in terms of rhythm matching.
Other detectable flaw was the transition between consecutive snippets.
Occasionally, the transition sounds somewhat unnatural, especially when performed between snippets that are part of different audio clips. This was addressed with fade in and fade out effects, but still, the transition can sometimes feel unpleasant. This could be addressed with interactive fitness, so that the user could provide the subjective evaluation missing in the automated system.

One of the main characteristics of Genetic Soundtracks is that even with the same configuration, i.e., the same video and set of audio clips, the system is capable of generating several distinct results.
We will address a particular configuration both in terms of analysis and resulting matchings.
The video used was a short 70\,s excerpt of \emph{Johan Hex} movie. In this excerpt, the first 44\,s show a conversation between still actors, and then there is a sudden transition to an action-packed shootout, that lasts until the end of the excerpt.
The audio clips used were two songs: (i)~Bittersweet by Apocalyptica, a very melodic song with cellos and voice, and drums in the chorus and (ii)~Fuel by Metallica, a very harsh thrash metal song.
In Fig.~\ref{fig:result}, we illustrate the result of processing both songs and the video.

Subjectively, one of the most interesting matchings performed consisted of using the first 44s of Bittersweet followed by Fuel from 149\,s to 175\,s. The matching is illustrated by the gray areas between graphics in Fig.~\ref{fig:result}.
The selected section of Bittersweet matches the video segment with smaller frame differencing values.
The video movement then increases for which the selected portion of Fuel constitutes a good choice for keeping up with the video.

Genetic Soundtracks has also shown to be versatile enough and produced diverse matchings.
For video scenes with higher frame differencing, distinct parts of the Fuel song were used.
Examples of segments employed are 126\,s to 143\,s, 193\,s to 212\,s, and 106\,s to 124\,s, all very similar in terms of noise and rythm.
Another result generated by the system consisted of using only Bittersweet for the soundtrack.
For the initial video scenes, those with less movement, the system used the segment from 0\,s to 50\,s, with only a cello and voice. For the following scenes, the soundtrack was composed by the segment from 151\,s to 171\,s, which corresponds to the chorus with drums.

\section{Conclusions and Future Work}

In this paper, we presented Genetic Soundtracks, an automated method for the matching of audio to a video.
Genetic Soundtracks adopts an automatic method for feature extraction.
The system extracts both the video movement and audio levels, which are used afterwards in the process of matching audio and video segments.
We demonstrated that genetic algorithms can be applied to the truncation and combination of audio segments, even from different clips.

The immediate follow-up work will include the extension of our adequacy concept to other synesthetic relations, through the extraction of distinct media features, or the use of meta-data embedded in the video and audio files.
Although Genetic Soundtracks may perform automatic matching of audio to video, presenting interesting results, we intend to empower the users by involving them in the creative edition of the soundtrack.
To this end, we will study innovative evolutionary combinations, to further explore the interactive definition of parameters and selection of individuals in the generations (interactive fitness)~\cite{videoArtEA}.
The users will therefore be able to subjectively judge different matchings of soundtracks to their videos, and guide the evolutionary search towards more personalized solutions with their human touch.

\bibliographystyle{IEEEtran}
\bibliography{biblio}

\end{document}